%% file: main.tex
\documentclass[12pt,a4paper]{article}
\usepackage{ifthen} % for conditional statements
\newboolean{pdflatex}
\setboolean{pdflatex}{true} % False for eps figures 

\newboolean{articletitles}
\setboolean{articletitles}{true} % False removes titles in references

\newboolean{uprightparticles}
\setboolean{uprightparticles}{false} %True for upright particle symbols

\def\paperauthors{LHCb collaboration} % Leave as is for PAPER, CONF and FIGURE
\def\paperasciititle{Observation of a resonant structure near the Ds+Ds- threshold in the B+ to Ds+ Ds- K+ decay} % Set ASCII title here !! MAKE sure it's only ASCII characters !! 
\def\papertitle{Observation of a resonant structure near the $\Dsp\Dsm$ threshold\\ in the \BuDsDsK decay} % Latex formatted title
\def\paperkeywords{{High Energy Physics}, {LHCb}} % Comma separated list
\def\papercopyright{\the\year\ CERN for the benefit of the LHCb collaboration} % new since 9/Apr/2018
\def\paperlicence{CC BY 4.0 licence}
\def\paperlicenceurl{https://creativecommons.org/licenses/by/4.0/}

\input{preamble}
\usepackage{longtable} % only for template; not usually to be used in PAPERs
\usepackage{rotating} % rotate table
\usepackage{multirow} % merge multiple rows

\begin{document}

%%%%%%%%%%%%%%%%%%%%%%%%%
%%%%% Title     %%%%%%%%%
%%%%%%%%%%%%%%%%%%%%%%%%%
\renewcommand{\thefootnote}{\fnsymbol{footnote}}
\setcounter{footnote}{1}

% %%%%%%% CHOOSE TITLE PAGE--------
%\onecolumn
%\input{title-LHCb-INT}
%\input{title-LHCb-ANA}
%\input{title-LHCb-CONF}
%\input{title-LHCb-FIGURE}
\input{title-LHCb-PAPER}

%\twocolumn
% %%%%%%%%%%%%% ---------

\renewcommand{\thefootnote}{\arabic{footnote}}
\setcounter{footnote}{0}

%%%%%%%%%%%%%%%%%%%%%%%%%%%%%%%%
%%%%%  Table of Content   %%%%%%
%%%%%%%%%%%%%%%%%%%%%%%%%%%%%%%%
%%%% Uncomment if desired
%\tableofcontents
\cleardoublepage

%%%%%%%%%%%%%%%%%%%%%%%%%
%%%%% Main text %%%%%%%%%
%%%%%%%%%%%%%%%%%%%%%%%%%

\pagestyle{plain} % restore page numbers for the main text
\setcounter{page}{1}
\pagenumbering{arabic}

%% Uncomment during review phase. 
%% Comment before a final submission.
%\linenumbers

%% This is the main body
%% It is useful to have a single file so comemnts are not missed in overleaf.

\input{body}

% Do not include this in any draft (just for information in the template)
%\input{acknowledgements_intro}
% Comment this in for paper drafts; do not include this in analysis note, conference and figure reports
\input{supplementary}

\input{acknowledgements}

%\input{appendix}

% This should be taken out in the final paper
%\input{supplementary}

%\input{LHCb_template}

\addcontentsline{toc}{section}{References}
%\setboolean{inbibliography}{true}
\bibliographystyle{LHCb}
\bibliography{main,standard,LHCb-PAPER,LHCb-CONF,LHCb-DP,LHCb-TDR}

\newpage
\input{Authorship_LHCb-PAPER-2022-018}

\clearpage

\end{document}

%% file: preamble.tex
\usepackage[top=1in, bottom=1.25in, left=1in, right=1in]{geometry}

\usepackage{microtype}
\usepackage{lineno}  % for line numbering during review
\usepackage{xspace} % To avoid problems with missing or double spaces after
\usepackage{caption} %these three command get the figure and table captions automatically small

\usepackage{graphicx}  % to include figures (can also use other packages)
\usepackage{color}
\usepackage{colortbl}
\graphicspath{{./figs/},{./eps/}} % Make Latex search fig subdir for figures
\usepackage{amsmath} % Adds a large collection of math symbols
\usepackage{amssymb}
\usepackage{amsfonts}
\usepackage{upgreek} % Adds in support for greek letters in roman typeset

\newcommand*\patchAmsMathEnvironmentForLineno[1]{%
\expandafter\let\csname old#1\expandafter\endcsname\csname #1\endcsname
\expandafter\let\csname oldend#1\expandafter\endcsname\csname
end#1\endcsname
 \renewenvironment{#1}%
   {\linenomath\csname old#1\endcsname}%
   {\csname oldend#1\endcsname\endlinenomath}%
}
\newcommand*\patchBothAmsMathEnvironmentsForLineno[1]{%
  \patchAmsMathEnvironmentForLineno{#1}%
  \patchAmsMathEnvironmentForLineno{#1*}%
}
\AtBeginDocument{%
\patchBothAmsMathEnvironmentsForLineno{equation}%
\patchBothAmsMathEnvironmentsForLineno{align}%
\patchBothAmsMathEnvironmentsForLineno{flalign}%
\patchBothAmsMathEnvironmentsForLineno{alignat}%
\patchBothAmsMathEnvironmentsForLineno{gather}%
\patchBothAmsMathEnvironmentsForLineno{multline}%
\patchBothAmsMathEnvironmentsForLineno{eqnarray}%
}

\usepackage{hyperxmp}

\usepackage[pdftex,
            pdfauthor={\paperauthors},
            pdftitle={\paperasciititle},
            pdfkeywords={\paperkeywords},
            pdfcopyright={Copyright (C) \papercopyright},
            pdflicenseurl={\paperlicenceurl}]{hyperref}
\usepackage[colorinlistoftodos,textsize=scriptsize]{todonotes}

\usepackage[bottom,flushmargin,hang,multiple]{footmisc}

\usepackage[all]{hypcap} % Internal hyperlinks to floats.

\input{lhcb-symbols-def} % Add in the predefined LHCb symbols
\input{mydefine}

\usepackage{cite} % Allows for ranges in citations
\usepackage{mciteplus}

%% file: lhcb-symbols-def.tex
\usepackage{xspace} 
\usepackage{upgreek}

\def\belle  {\mbox{Belle}\xspace}

\def\MagUp {\mbox{\em Mag\kern -0.05em Up}\xspace}

\ifthenelse{\boolean{uprightparticles}}%
{

 \def\Ppi         {\ensuremath{\uppi}\xspace}

 \def\Pchi        {\ensuremath{\upchi}\xspace}                 
 \def\Ppsi        {\ensuremath{\uppsi}\xspace}

 \def\PDelta      {\ensuremath{\Delta}\xspace}                 
 \def\PXi         {\ensuremath{\Xi}\xspace}                 
 \def\PLambda     {\ensuremath{\Lambda}\xspace}                 
 \def\PSigma      {\ensuremath{\Sigma}\xspace}                 
 \def\POmega      {\ensuremath{\Omega}\xspace}                 
 \def\PUpsilon    {\ensuremath{\Upsilon}\xspace}
 \let\oldPi\Pi
 \def\PPi         {\ensuremath{\oldPi}\xspace}

 \def\PB      {\ensuremath{\mathrm{B}}\xspace}                 
                  
 \def\PD      {\ensuremath{\mathrm{D}}\xspace}

 \def\PJ      {\ensuremath{\mathrm{J}}\xspace}                 
 \def\PK      {\ensuremath{\mathrm{K}}\xspace}

 \def\Pc      {\ensuremath{\mathrm{c}}\xspace}                 
 \def\Pd      {\ensuremath{\mathrm{d}}\xspace}

 \def\Pi      {\ensuremath{\mathrm{i}}\xspace}

 \def\Pq      {\ensuremath{\mathrm{q}}\xspace}                 
                  
 \def\Ps      {\ensuremath{\mathrm{s}}\xspace}                 
                  
 \def\Pu      {\ensuremath{\mathrm{u}}\xspace}

 \def\thebaroffset{0.0em}
}
{

 \def\Ppi         {\ensuremath{\pi}\xspace}

 \def\Pchi        {\ensuremath{\chi}\xspace}                 
 \def\Ppsi        {\ensuremath{\psi}\xspace}                 
                  
 \mathchardef\PDelta="7101
 \mathchardef\PXi="7104
 \mathchardef\PLambda="7103
 \mathchardef\PSigma="7106
 \mathchardef\POmega="710A
 \mathchardef\PUpsilon="7107
 \mathchardef\PPi="7105
                  
 \def\PB      {\ensuremath{B}\xspace}                 
                  
 \def\PD      {\ensuremath{D}\xspace}

 \def\PJ      {\ensuremath{J}\xspace}                 
 \def\PK      {\ensuremath{K}\xspace}

 \def\Pc      {\ensuremath{c}\xspace}                 
 \def\Pd      {\ensuremath{d}\xspace}

 \def\Pi      {\ensuremath{i}\xspace}

 \def\Pq      {\ensuremath{q}\xspace}                 
                  
 \def\Ps      {\ensuremath{s}\xspace}                 
                  
 \def\Pu      {\ensuremath{u}\xspace}

 \def\thebaroffset{0.18em}
}
\newcommand{\offsetoverline}[2][\thebaroffset]{\kern #1\overline{\kern -#1 #2}}%

\makeatletter
\ifcase \@ptsize \relax% 10pt
  \newcommand{\miniscule}{\@setfontsize\miniscule{4}{5}}% \tiny: 5/6
\or% 11pt
  \newcommand{\miniscule}{\@setfontsize\miniscule{5}{6}}% \tiny: 6/7
\or% 12pt
  \newcommand{\miniscule}{\@setfontsize\miniscule{5}{6}}% \tiny: 6/7
\fi
\makeatother

\DeclareRobustCommand{\optbar}[1]{\shortstack{{\miniscule (\rule[.5ex]{1.25em}{.18mm})}
  \\ [-.7ex] $#1$}}

\def\quark     {{\ensuremath{\Pq}}\xspace}
\def\quarkbar  {{\ensuremath{\overline \quark}}\xspace}

\def\uquark    {{\ensuremath{\Pu}}\xspace}
\def\uquarkbar {{\ensuremath{\overline \uquark}}\xspace}

\def\dquark    {{\ensuremath{\Pd}}\xspace}
\def\dquarkbar {{\ensuremath{\overline \dquark}}\xspace}

\def\squark    {{\ensuremath{\Ps}}\xspace}
\def\squarkbar {{\ensuremath{\overline \squark}}\xspace}

\def\cquark    {{\ensuremath{\Pc}}\xspace}
\def\cquarkbar {{\ensuremath{\overline \cquark}}\xspace}

\def\pion   {{\ensuremath{\Ppi}}\xspace}

\def\pip    {{\ensuremath{\pion^+}}\xspace}

\def\kaon    {{\ensuremath{\PK}}\xspace}
\def\KorKbar {\kern \thebaroffset\optbar{\kern -\thebaroffset \PK}{}\xspace}

\def\Kp      {{\ensuremath{\kaon^+}}\xspace}
\def\Km      {{\ensuremath{\kaon^-}}\xspace}

\def\Dbar    {{\ensuremath{\offsetoverline{\PD}}}\xspace}
\def\D       {{\ensuremath{\PD}}\xspace}

\def\DorDbar {\kern \thebaroffset\optbar{\kern -\thebaroffset \PD}\xspace}

\def\Dp      {{\ensuremath{\D^+}}\xspace}
\def\Dm      {{\ensuremath{\D^-}}\xspace}

\def\DpDm    {\ensuremath{\Dp {\kern -0.16em \Dm}}\xspace}

\def\Ds      {{\ensuremath{\D^+_\squark}}\xspace}
\def\Dsp     {{\ensuremath{\D^+_\squark}}\xspace}
\def\Dsm     {{\ensuremath{\D^-_\squark}}\xspace}

\def\B       {{\ensuremath{\PB}}\xspace}

\def\BorBbar {\kern \thebaroffset\optbar{\kern -\thebaroffset \PB}\xspace}

\def\Bd      {{\ensuremath{\B^0}}\xspace}

\def\BdorBdbar {\kern \thebaroffset\optbar{\kern -\thebaroffset \Bd}\xspace}
\def\Bu      {{\ensuremath{\B^+}}\xspace}

\def\Bp      {{\ensuremath{\Bu}}\xspace}

\def\Bs      {{\ensuremath{\B^0_\squark}}\xspace}

\def\BsorBsbar {\kern \thebaroffset\optbar{\kern -\thebaroffset \Bs}\xspace}

\def\jpsi     {{\ensuremath{{\PJ\mskip -3mu/\mskip -2mu\Ppsi}}}\xspace}

\def\chiczero {{\ensuremath{\Pchi_{\cquark 0}}}\xspace}

\def\chictwo  {{\ensuremath{\Pchi_{\cquark 2}}}\xspace}

\def\Y#1S{\ensuremath{\PUpsilon{(#1S)}}\xspace}

\def\LorLbar     {\kern \thebaroffset\optbar{\kern -\thebaroffset \PLambda}\xspace}

\newcommand{\decay}[2]{\ensuremath{#1\!\to #2}\xspace} 

\def\to                 {\ensuremath{\rightarrow}\xspace}

\def\AT#1     {\ensuremath{A_{\mathrm{T}}^{#1}}\xspace}           % 2

\def\C#1      {\ensuremath{\mathcal{C}_{#1}}\xspace}                       % 9
\def\Cp#1     {\ensuremath{\mathcal{C}_{#1}^{'}}\xspace}                    % 7
\def\Ceff#1   {\ensuremath{\mathcal{C}_{#1}^{\mathrm{(eff)}}}\xspace}        % 9  
\def\Cpeff#1  {\ensuremath{\mathcal{C}_{#1}^{'\mathrm{(eff)}}}\xspace}       % 7
\def\Ope#1    {\ensuremath{\mathcal{O}_{#1}}\xspace}                       % 2
\def\Opep#1   {\ensuremath{\mathcal{O}_{#1}^{'}}\xspace}                    % 7

\newcommand{\aunit}[1]{\ensuremath{\text{\,#1}}}       
\newcommand{\tev}{\aunit{Te\kern -0.1em V}\xspace}
\newcommand{\gev}{\aunit{Ge\kern -0.1em V}\xspace}
\newcommand{\mev}{\aunit{Me\kern -0.1em V}\xspace}
\newcommand{\kev}{\aunit{ke\kern -0.1em V}\xspace}
\newcommand{\ev}{\aunit{e\kern -0.1em V}\xspace}
 
\newcommand{\mevc}{\ensuremath{\aunit{Me\kern -0.1em V\!/}c}\xspace}
\newcommand{\gevc}{\ensuremath{\aunit{Ge\kern -0.1em V\!/}c}\xspace}
\newcommand{\mevcc}{\ensuremath{\aunit{Me\kern -0.1em V\!/}c^2}\xspace}
\newcommand{\gevcc}{\ensuremath{\aunit{Ge\kern -0.1em V\!/}c^2}\xspace}
\def\fm   {\aunit{fm}\xspace}

\def\fb   {\ensuremath{\aunit{fb}}\xspace}
\def\invfb   {\ensuremath{\fb^{-1}}\xspace}

\def\gsim{{~\raise.15em\hbox{$>$}\kern-.85em
          \lower.35em\hbox{$\sim$}~}\xspace}
\def\lsim{{~\raise.15em\hbox{$<$}\kern-.85em
          \lower.35em\hbox{$\sim$}~}\xspace}

\def\sPlot{\mbox{\em sPlot}\xspace}
\def\sFit{\mbox{\em sFit}\xspace}
\def\tell1  {TELL1\xspace}
\def\ukl1   {UKL1\xspace}

\newcommand{\ie}{\mbox{\itshape i.e.}\xspace}
\newcommand{\etal}{\mbox{\itshape et al.}\xspace}

\newcommand{\lhcborcid}[1]{\href{https://orcid.org/#1}{\hspace*{0.1em}\raisebox{-0.45ex}{\includegraphics[width=1em]{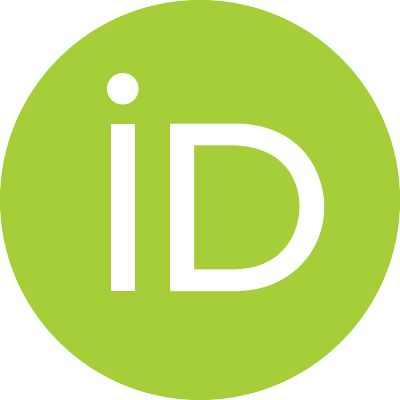}}}}

%% file: mydefine.tex
\def\BuDsDsK       {\ensuremath{\mbox{\decay{\Bp}{\Dsp\Dsm\Kp}}}\xspace}
\def\BuDDK         {\ensuremath{\mbox{\decay{\Bp}{\Dp\Dm\Kp}}}\xspace}

\def\DsKKpi        {\ensuremath{\mbox{\Ds\to\Km\Kp\pip}}\xspace}
\def\DKpipi        {\ensuremath{\mbox{\Dp\to\Km\pip\pip}}\xspace}

\def\mDsDs        {\ensuremath{\mbox{$m(\Dsp\Dsm)$}}\xspace}
\def\mDsK        {\ensuremath{\mbox{$m(\Dsm\Kp)$}}\xspace}

\def\BrDs          {\ensuremath{\mbox{${\cal B}$(\DsKKpi)}}\xspace}

\def\BrD          {\ensuremath{\mbox{${\cal B}$(\DKpipi)}}\xspace}

\def\Like        {\ensuremath{{\cal L}}\xspace}

%% file: title-LHCb-PAPER.tex
% ===============================================================================
% Purpose: LHCb-PAPER journal paper title page template
% Author: 
% Created on: 2010-09-25
% ===============================================================================

%%%%%%%%%%%%%%%%%%%%%%%%%
%%%%%  TITLE PAGE  %%%%%%
%%%%%%%%%%%%%%%%%%%%%%%%%
\begin{titlepage}
\pagenumbering{roman}

% Header ---------------------------------------------------
\vspace*{-1.5cm}
\centerline{\large EUROPEAN ORGANIZATION FOR NUCLEAR RESEARCH (CERN)}
\vspace*{1.5cm}
\noindent
\begin{tabular*}{\linewidth}{lc@{\extracolsep{\fill}}r@{\extracolsep{0pt}}}
\ifthenelse{\boolean{pdflatex}}% Logo format choice
{\vspace*{-1.5cm}\mbox{\!\!\!\includegraphics[width=.14\textwidth]{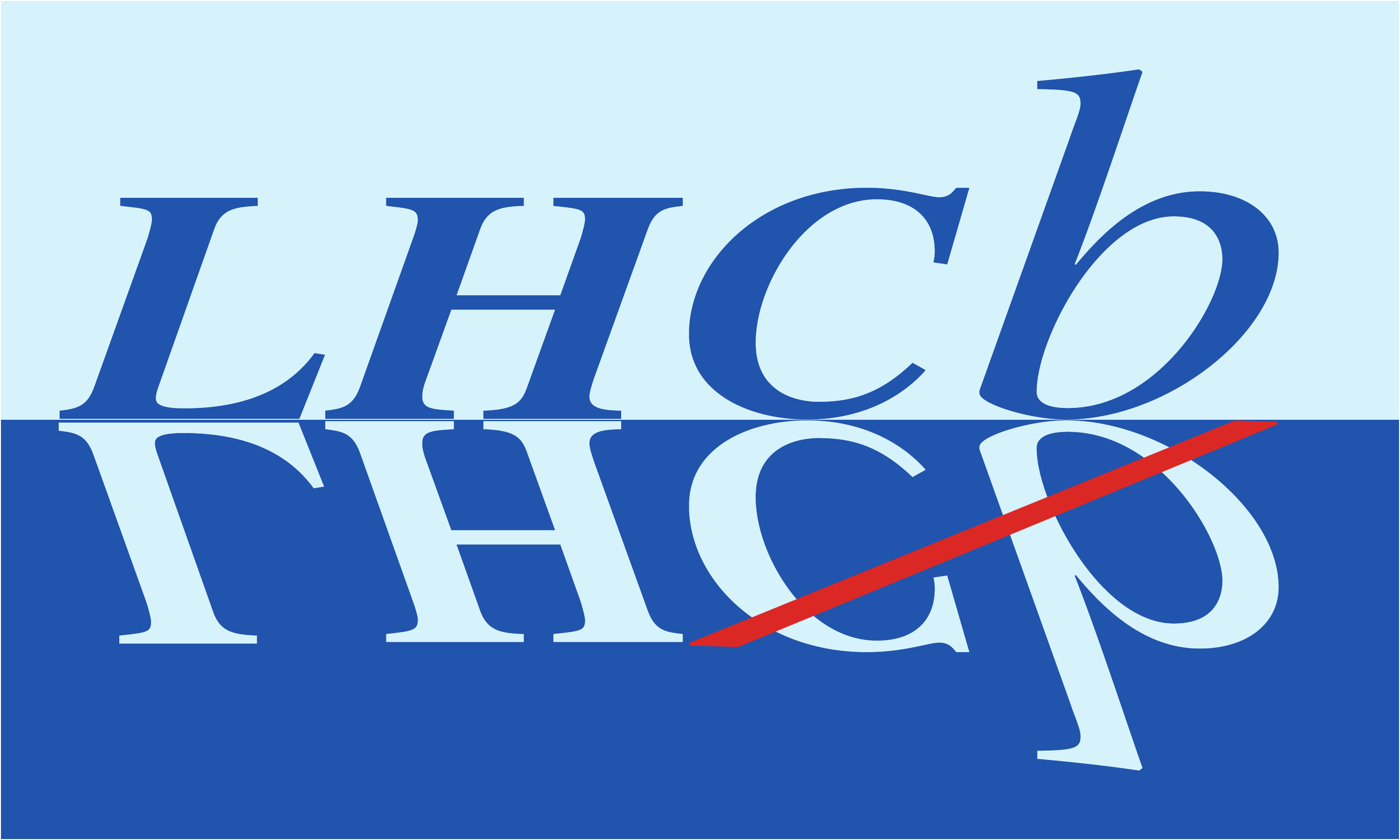}} & &}%
{\vspace*{-1.2cm}\mbox{\!\!\!\includegraphics[width=.12\textwidth]{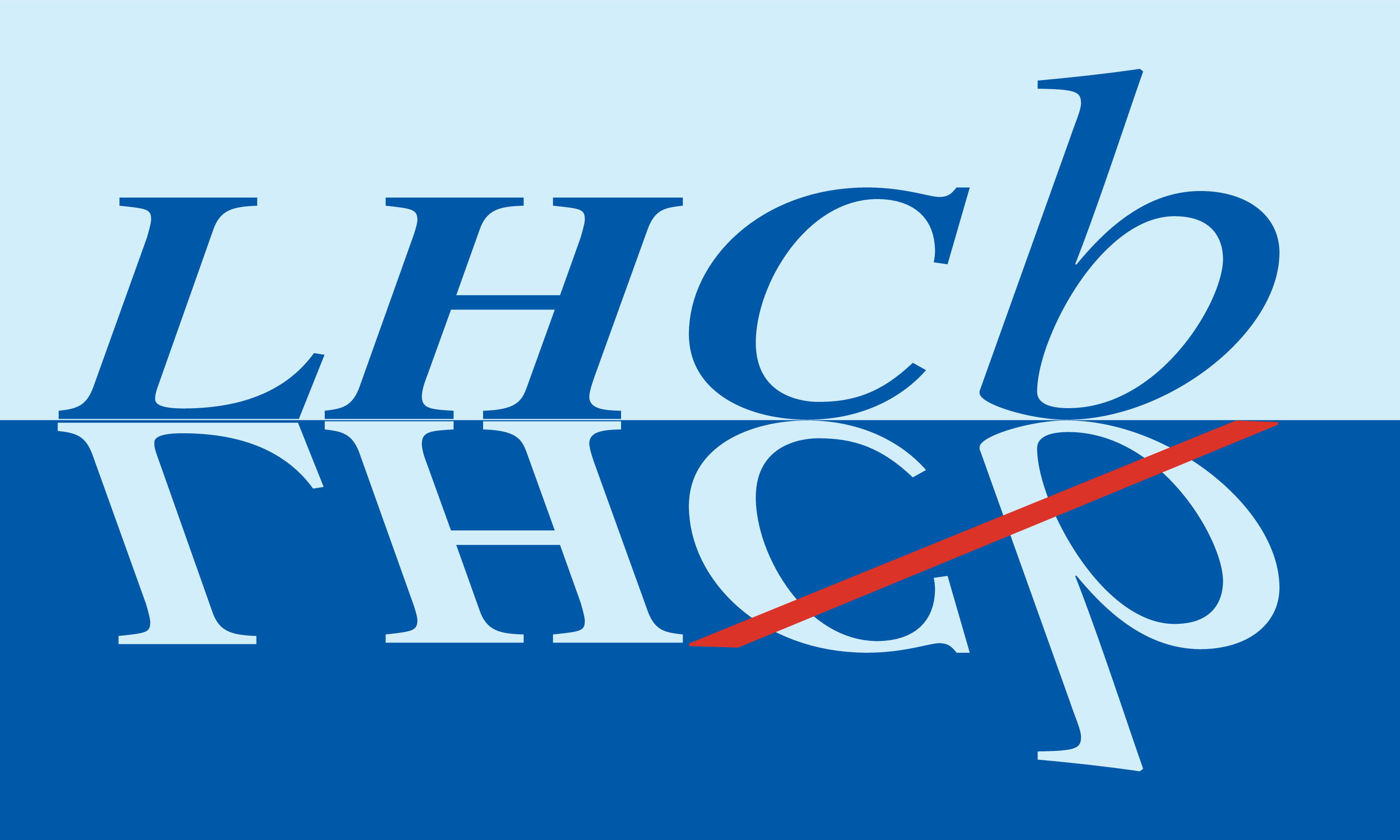}} & &}%
\\
 & & CERN-EP-2022-200 \\  % ID 
 & & LHCb-PAPER-2022-018 \\  % ID 
 & & \today \\ % Date - Can also hardwire e.g.: 23 March 2010
 & & \\
% not in paper \hline
\end{tabular*}

\vspace*{4.0cm}

% Title --------------------------------------------------
{\normalfont\bfseries\boldmath\huge
\begin{center}
% DO NOT EDIT HERE. Instead edit macro in main.tex to keep metadata correct
  \papertitle 
\end{center}
}

\vspace*{2.0cm}

% Authors -------------------------------------------------
\begin{center}
%In the footnote, replace 'paper' by 'Letter' in case of submission to PRL or PLB 
% Edit macro in main.tex to keep metadata correct
\paperauthors\footnote{Authors are listed at the end of this paper.}
\end{center}

\vspace{\fill}

% Abstract -----------------------------------------------
\begin{abstract}
  \noindent
 An amplitude analysis of the \BuDsDsK decay is carried out to study for the first time its intermediate resonant contributions, using proton-proton collision data collected with the LHCb detector at centre-of-mass energies of 7, 8 and 13\tev. A near-threshold peaking structure, referred to as $X(3960)$, is observed in the \Dsp\Dsm invariant-mass spectrum with significance greater than 12 standard deviations.
The mass, width and the quantum numbers of the structure are measured to be $3956\pm5\pm10$\mev, $43\pm13\pm8$\mev and $J^{PC}=0^{++}$,  respectively, where the first uncertainties are statistical and the second systematic.
 The properties of the new structure are consistent with recent theoretical predictions for a state composed of \cquark\cquarkbar\squark\squarkbar quarks.
 Evidence for an additional structure  is found around 4140~\mev in the \Dsp\Dsm invariant mass, which might be caused either by a new resonance with the $0^{++}$ assignment or by a $\jpsi\phi\leftrightarrow\Dsp\Dsm$ coupled-channel effect. 
  
\end{abstract}

\vspace*{2.0cm}

\begin{center}
  Published in
%   JHEP /
%   Phys.~Rev.~D /
  Phys.~Rev.~Lett. {\bf 131} (2023) 071901. 
%  Phys.~Lett.~B /
%   Eur.~Phys.~J.~C /
%   %  Nucl.~Phys.~B /
%   Chin.~Phys.~C /
%   Nature~Physics /
%   sciPost~Physics /
%   J. Instr. /
%   Instruments 
\end{center}

\vspace{\fill}

{\footnotesize 
% Edit macro in main.tex to keep metadata correct
\centerline{\copyright~\papercopyright. \href{\paperlicenceurl}{\paperlicence}.}}
\vspace*{2mm}

\end{titlepage}

%%%%%%%%%%%%%%%%%%%%%%%%%%%%%%%%
%%%%%  EOD OF TITLE PAGE  %%%%%%
%%%%%%%%%%%%%%%%%%%%%%%%%%%%%%%%

%  empty page follows the title page ----
\newpage
\setcounter{page}{2}
\mbox{~}
%\newpage
%
%% Author List ----------------------------
%%  You need to get a new author list!
%\input{LHCb_authorlist.tex}
%
%The author list for journal publications is provided by the Membership Committee shortly after 'approval to go to paper' has been given.
%%It will be made available on the page
%%\verb!http://www.physik.uzh.ch/~strauman/forMemCo/LHCb-PAPER-XXXX-XXX/! .
%It will be sent to you by email shortly after a paper number has beens assigned.
%The author list should be included already at first circulation, 
%to allow new members of the collaboration to verify whether they have been included correctly.
%Occasionally a misspelled name is corrected or associated institutions become full members.
%In that case, a new author list will be sent to you.
%In case line numbering doesn't work well after including the authorlist, try moving the \verb!\bigskip! after the last author to a separate line.
%
%
%The authorship for Conference Reports should be ``The LHCb
%  collaboration'', with a footnote giving the name(s) of the contact
%  author(s), but without the full list of collaboration names.

%% file: body.tex
%\section{Introduction}
%\label{sec:Introduction}

%composed of other than a quark-antiquark pair and three quarks
Exotic hadrons\footnote{Hadrons that are not composed either of a quark-antiquark pair or of three quarks or three antiquarks are collectively called exotic hadrons.}
play a crucial role in studies of Quantum Chromodynamics~(QCD), and provide a unique window to understand the nature of the strong interaction. %the strong force. 
Dozens of charged states with hidden charm or beauty,  which imply exotic nature,  such as $Z_c(4430)^+$~\cite{Belle4430,Belle4430Jp}, $Z_b(10610)^+$~\cite{BelleZb}, $Z_c(3900)^+$~\cite{BESIII3900,Belle3900,CLOE3900},
$Z_c(4020)^+$~\cite{BESIII:2013mhi,BESIII:2013ouc},
$P_c(4450)^+$~\cite{LHCb-PAPER-2015-029,LHCb-PAPER-2019-014},
$Z_{cs}(3985)^+$~\cite{BESIIIZcs3985}, $Z_{cs}(4000)^+$~\cite{LHCb-PAPER-2020-044}, have been recently discovered by various experiments.~\footnote{The inclusion of charge-conjugate processes is always implied and natural units with $\hbar=c=1$ are used throughout the Letter.}
Over the last two years, the LHCb collaboration reported three new open-charm tetraquark states, $X_{0,1}(2900)^0$~\cite{LHCb-PAPER-2020-024,LHCb-PAPER-2020-025} and $T_{cc}(3875)^+$~\cite{LHCb-PAPER-2021-031,LHCb-PAPER-2021-032}, composed of $\cquark\squark\uquarkbar\dquarkbar$ and $\cquark\cquark\uquarkbar\dquarkbar$ quarks, respectively.
Interestingly, most of these states have masses close to thresholds of hadron pairs, which may indicate that they are hadronic molecules loosely bound by deuteron-like meson-exchange forces~\cite{molecule:2016hlt,molecule:2016jzb,molecule:6900,molecule:2900}. There are a number of other possible explanations, including that these particles are compact multi-quark states~\cite{tetraquark:2013lda,tetraquark:2013xfa,tetraquark:2004vq}, hadroquarkonia in which a $c\bar{c}$ core is bound to light quarks and/or gluons via chromo-electric dipole interactions~\cite{Dubynskiy:2008di,hadrocc:2016dru}, or cusps produced by near-threshold 
kinematics involving open-charm hadrons, or other dominant processes~\cite{cusp:2011jr,cusp:2019twa}.

The $\chiczero(3930)$ state was observed by the LHCb collaboration in the $D^+D^-$ invariant-mass spectrum~\cite{LHCb-PAPER-2020-025}.
The mass and width of this state are consistent with those of the $X(3915)$ resonance observed in the $\omega \jpsi$ invariant-mass spectrum~\cite{3915:belle,3915:babar,gg3915:belle,gg3915:babar}. Moreover, the $X(3915)$ has preferred spin ($J$), parity ($P$), and charge-parity ($C$) quantum numbers of $J^{PC}=0^{++}$~\cite{gg3915:babar,PDG2020}, 
so the two states are treated as a single hadron in the following discussions unless otherwise specified.
However, the $\chi_{c0}(3930)$ state is not considered to be consistent with being a candidate for either the $\chiczero(2P)$ or $\chiczero(3P)$ state~\cite{Barnes:2005pb,Radford:2007vd,cc:KTChao,3Pj:Ping,3860:Guo}.
Lebed \etal~\cite{3915:Lebed} propose that it is the lightest $c\bar{c} s \bar{s}$ state. Calculations based on QCD sum rules~\cite{3915:2017dpy} favour the $\chi_{c0}(3930)$ state as a $0^{++}$ $[\cquark\quark][\cquarkbar\quarkbar]$ (where $\quark=\uquark,\dquark$) 
or $[\cquark\squark][\cquarkbar\squarkbar]$ tetraquark. Recent lattice QCD results also indicate that this state  is dominated by the $c\bar{c} s \bar{s}$ constituents~\cite{3915:LQCD}.
The \Dsp\Dsm molecular interpretation is also possible, as suggested by the quark delocalization color-screening model~\cite{3915:jlping} and other phenomenological studies~\cite{Meng:2020cbk,Dong:2021juy}.
All these developments point to a potential resonant structure in the vicinity of the threshold in the \Dsp\Dsm invariant-mass spectrum.

Previously, only the Belle experiment studied the $\Dsp\Dsm$ invariant-mass spectrum in processes involving initial-state radiation,
where only $1^{--}$ charmonium(-like) states can contribute~\cite{Belle:2010fwv}. The \BuDsDsK process,  given its large branching fraction measured in the accompanying paper~\cite{LHCb-PAPER-2022-019}, provides a good opportunity to study resonances in the $\Dsp\Dsm$ final states, both scalars and those of higher spin, such as the $0^{++}$ \mbox{charmonium(-like)} states $\chiczero(4500)$ and $\chiczero(4700)$ possibly having an intrinsic $\cquark\cquarkbar\squark\squarkbar$ component~\cite{LHCb-PAPER-2020-044}, the well-known $1^{--}$ charmonium states, such as $\psi(4040)$, $\psi(4160)$, $\psi(4260)$, $\psi(4415)$ and $\psi(4660)$~\cite{PDG2020,4260:zheng,4660:zheng}.

In this Letter, an amplitude analysis of about 360 reconstructed \BuDsDsK signal decays is presented, leading to the first observation of a near-threshold  peaking structure in the \Dsp\Dsm system, denoted by $X(3960)$.
The analysis is based on proton-proton ($pp$) collision data collected by the LHCb experiment at centre-of-mass energies of 7, 8 and 13\tev between 2011 and 2018,
corresponding to an integrated luminosity of 9\invfb. 
The \Dsp candidates are reconstructed via the $\Dsp\to\Km\Kp\pip$ decay.
The details of the detector, data and simulation, selection criteria, background composition and \Bp invariant-mass fit can be found in the accompanying paper~\cite{LHCb-PAPER-2022-019}.  

\begin{figure}[!tbp]
    \centering
    \includegraphics[width=0.6\textwidth]{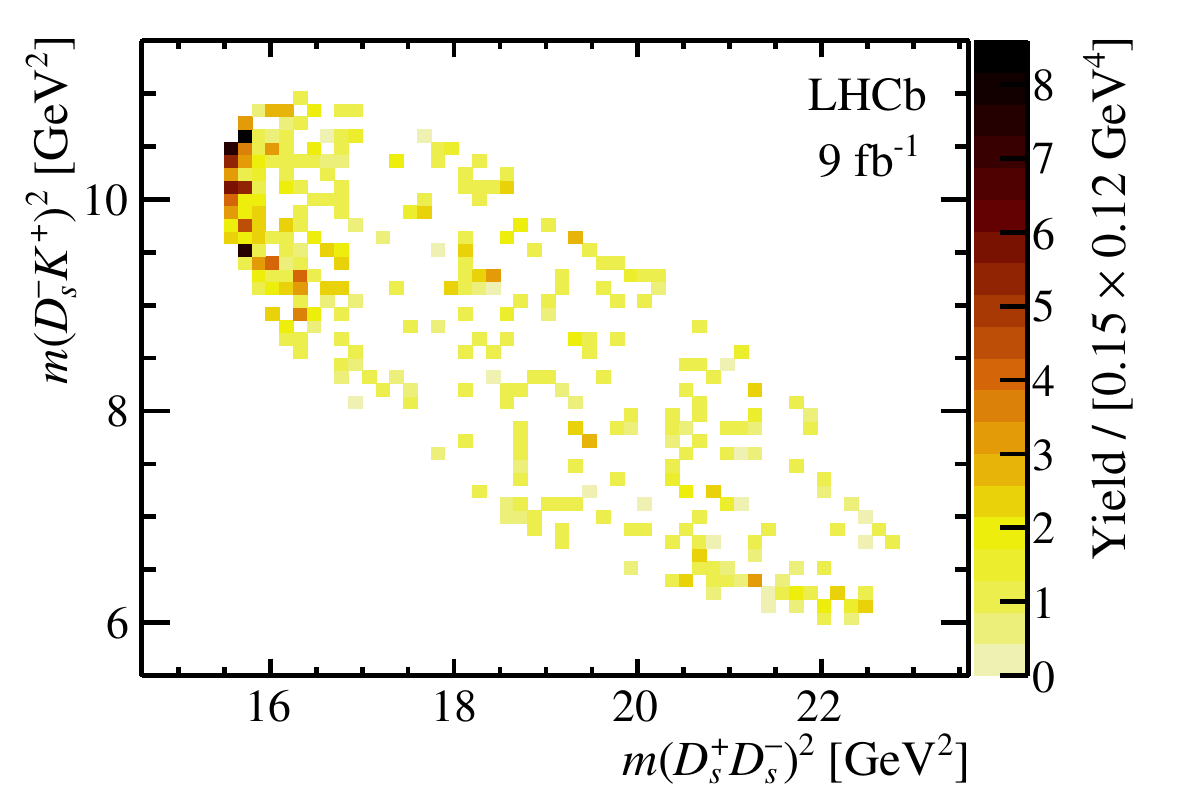}

    \caption{Dalitz-plot distribution for the \BuDsDsK decay after background subtraction. 
    }
    \label{fig:dlz_sig}
\end{figure}

To improve the resolution on the masses of the two-body combinations that are used in the amplitude analysis, the four momentum of each final-state particle is determined from a kinematic fit~\cite{Hulsbergen:2005pu} where the \Bp mass is constrained to its known value~\cite{PDG2020}.
Figure~\ref{fig:dlz_sig} shows the resulting Dalitz-plot distribution for the \BuDsDsK signal decays, %as a function of the \Dsp\Dsm and \Dsm\Kp invariant masses squared, 
where the non-\Bp background is subtracted by the \sPlot technique~\cite{Pivk:2004ty} with the reconstructed $\Bp$ mass as the discriminating variable. 
The most evident feature is the band near the \Dsp\Dsm threshold.
To validate that this peaking structure is not due to the combinatorial background, the \Dsp\Dsm invariant-mass distribution of candidates in the $\Bp$ mass region from 5360 to 5600\mev is investigated and no peak is observed.

\begin{figure}[bt]
\centering
\hspace{-0.8cm}
\includegraphics[width=0.495\textwidth]{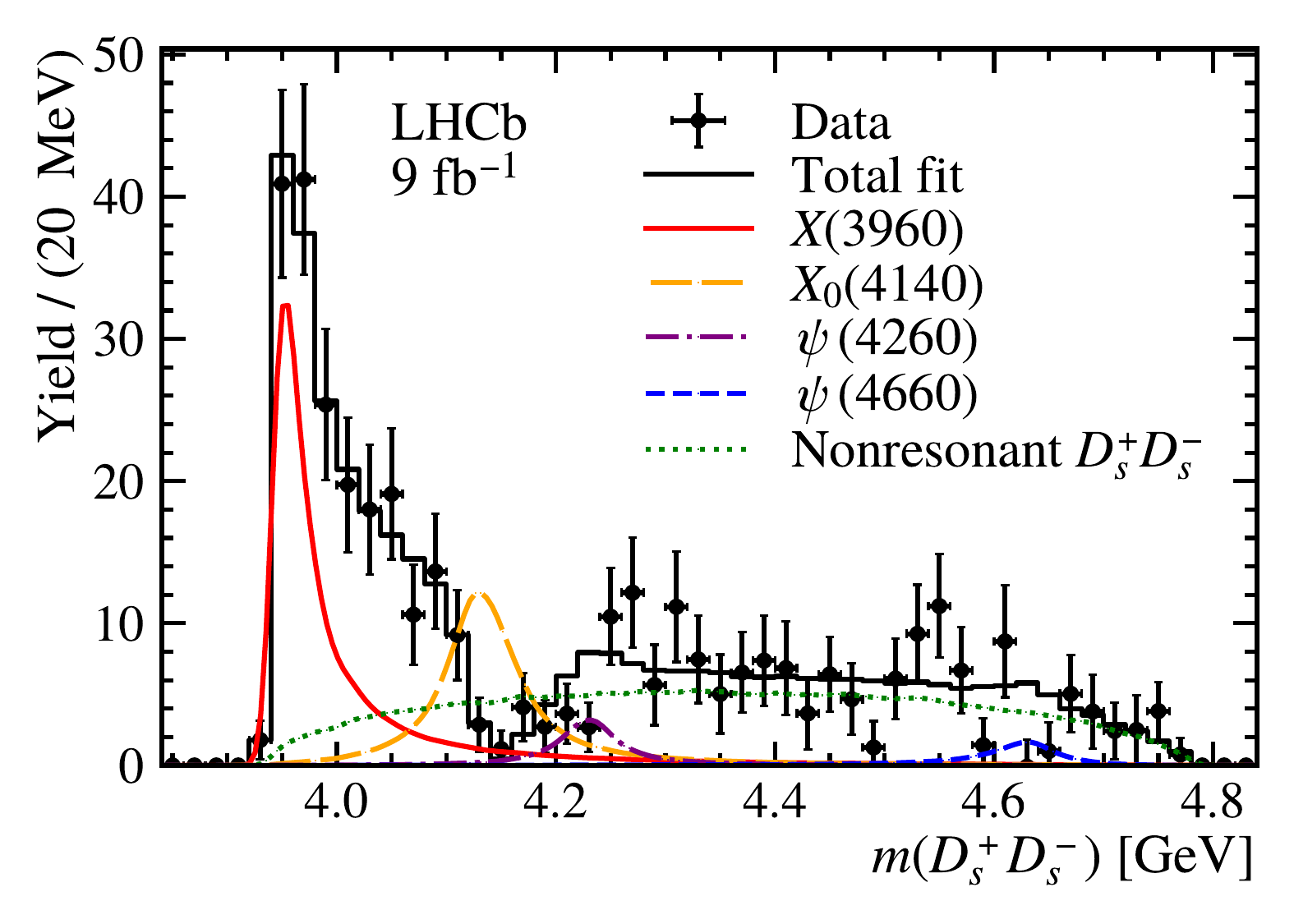} 
\includegraphics[width=0.495\textwidth]{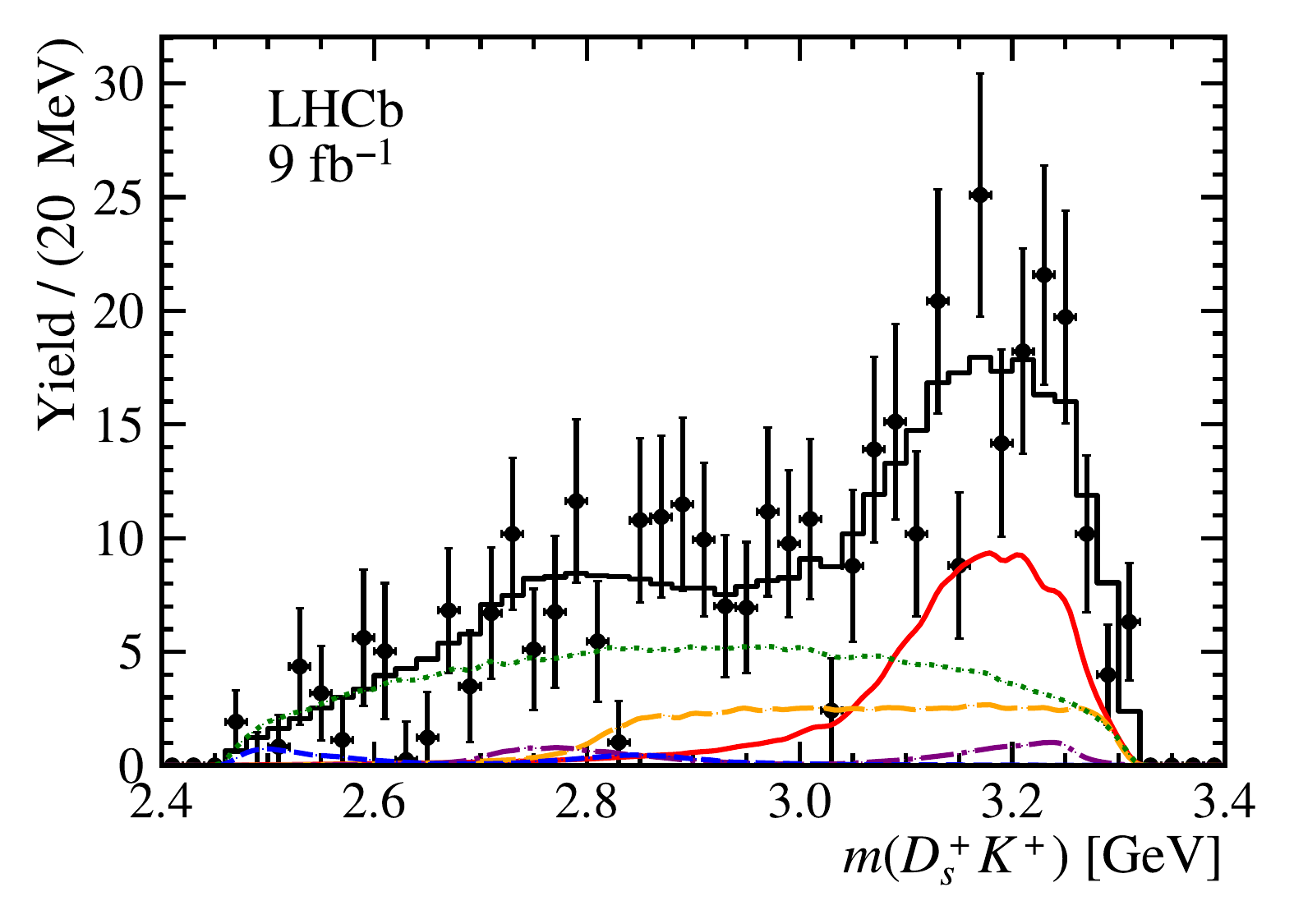}
\includegraphics[width=0.495\textwidth]{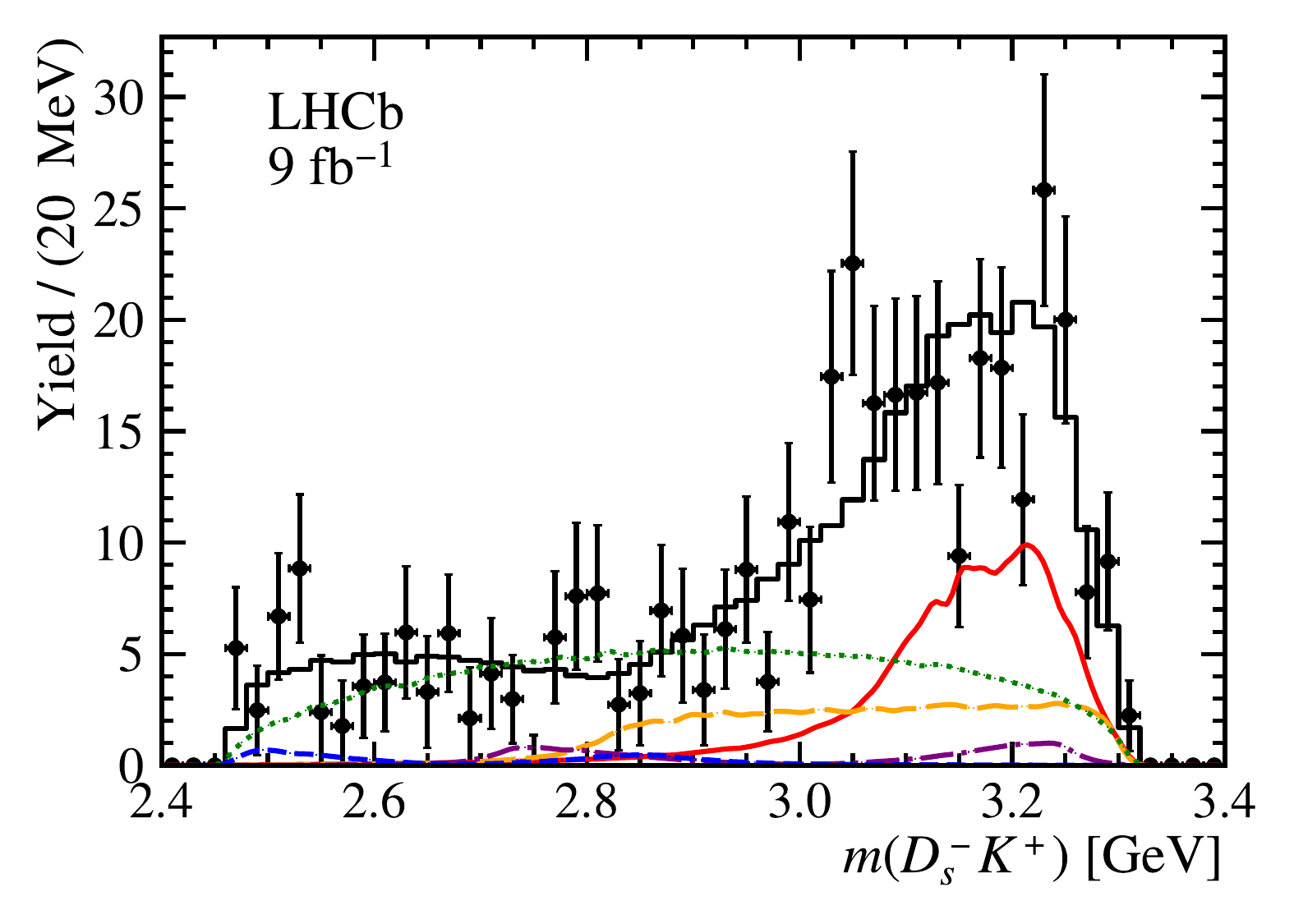}
\caption{Background-subtracted invariant-mass distributions (top left) \mDsDs, (top right) $m(\Dsp\Kp)$ and (bottom) \mDsK for the \BuDsDsK  signal. 
The projections of the fit with the baseline amplitude model are also shown.
\label{fig:dlz_fit}} 
\end{figure}

Employing an unbinned maximum-likelihood method, an amplitude fit with the \sFit technique~\cite{Xie:2009rka} is performed to investigate the intermediate states and determine the quantum numbers $J^{PC}$ of any new particle.
Two known $1^{--}$ charmonium states,  $\psi(4260)$ and $\psi(4660)$~\cite{PDG2020,4260:zheng,4660:zheng}, and two new $0^{++}$ $X$ states are needed to fit the structures in the \Dsp\Dsm spectrum. One of these scalars, $X(3960)$, describes the \Dsp\Dsm threshold enhancement and the other, designated $X_0(4140)$, is necessary to model the dip around 4140\mev, as shown in Fig.~\ref{fig:dlz_fit}. The subscript 0 is used to distinguish the latter from the $1^{++}$ $X(4140)$ state seen in the $\jpsi\phi$ final state~\cite{PDG2020}.  
Additionally, an $S$-wave three-body phase-space function~\cite{PDG2020} is employed to model the nonresonant~(NR) \BuDsDsK component. Since no significant contribution of any state is observed in either the \Dsm\Kp or \Dsp\Kp systems, these five contributions constitute the baseline model. 

The helicity formalism~\cite{Chung:1971ri,*Richman:1984gh,*Jacob:1959at} is used to construct the amplitude model of the \BuDsDsK decay, with a similar approach applied to previous LHCb analyses of \Bp and \Bs decays to three pseudoscalar particles~\cite{LHCb-PAPER-2014-036,LHCb-PAPER-2016-026,LHCb-PAPER-2019-017,LHCb-PAPER-2020-025}. 
The resonant structure near the \Dsp\Dsm mass threshold is parameterised by a Flatt\'{e}-like function~\cite{PDG2020,flatte:1976xu,molecule:6900} depending on the invariant mass $m$
\begin{equation}
\label{eq:flatte}
    R\left(m \mid M_{0}, g_{j}\right)=\frac{1}{M_{0}^{2}-m^{2}-i M_{0} \sum_{j} g_{j} \rho_{j}(m)},
\end{equation}
where $M_0$ is the mass of the resonance, $g_{j}$ denotes the coupling of this resonance to the $j$-th channel,
$\rho_j(m)$ is the phase-space factor~\cite{PDG2020} for the $j$-th two-body decay.
When the value of $m$ is below the threshold of the channel $j$, \ie $q_j^2<0$, an analytic continuation is applied for $q_j = i\sqrt{-q_j^2}$~\cite{flatte:1976xu,kmatrix:2002}. 
 The total width of the resonance is calculated as $\Gamma_0=\sum_{j} g_{j} \rho_{j}(M_0)$. 
In the baseline model, only the \Dsp\Dsm channel ($j=1$) is included in the Flatt\'{e}-like parameterisation.

Other resonances are modelled by a relativistic Breit--Wigner function ${\rm BW}\!\left(m  \mid M_{0}, \Gamma_{0} \right)$ with a mass-dependent width~\cite{PDG2020}.
The radius of each resonance entering the Blatt--Weisskopf barrier factor~\cite{Blatt:1952ije,VonHippel:1972fg,Chung:1995dx} is set to $3\gev^{-1}$, corresponding to about $0.6\fm$.

The total probability density function is the squared modulus of the total decay amplitude multiplied by the efficiency,
normalised to ensure that the integral over the Dalitz plot is unity. 
 The fit fraction $\mathcal{F}_i$ expresses the fraction of the total rate due to the component $i$, and the interference fraction $\mathcal{I}_{ij}$ describes the interference between components $i$ and $j$. They are defined in Eqs.~(18) and (19) of Ref.~\cite{LHCb-PAPER-2016-026}, such that $\sum_i {\cal F}_i + \sum_{i<j} \mathcal{I}_{ij} = 1$.

As shown in Fig.~\ref{fig:dlz_fit}, the two-body mass distributions are well modelled by the baseline amplitude fit. The corresponding numerical results are summarised in Table~\ref{tab:conclusion}, including the mass, width, fit fraction, and significance ($\mathcal{S}$) of each component. 
The significance of a given component is evaluated by assuming that the change of twice the negative log-likelihood~($-2\ln\Like$) between the baseline fit and the fit without that component obeys a $\chi^2$ distribution, where the number of degrees of freedom~(n.d.f.) 
is given by the change in the number
of free parameters.  All the components included in the baseline model have a statistical significance greater than three standard deviations~($\sigma$), where the $X(3960)$ and $X_0(4140)$ states are found to be 14.6\,$\sigma$ and 4.1\,$\sigma$ significant, respectively.
The obtained significances for the $X(3960)$ and $X_0(4140)$ resonances are also validated using pseudoexperiments.

\begin{table}[btp]
    \centering
    \renewcommand\arraystretch{1.1} 
    \caption{Summary of the main results obtained with the baseline model, where the first uncertainty is statistical and the second systematic. The last column shows the signal significance with~(without) the systematic uncertainty included.}
    \label{tab:conclusion}
    \begin{tabular}{lccccc}\hline
     Component & $J^{PC}$ & $M_0$~(MeV) & $\Gamma_0$~(MeV) & $\mathcal{F}$ (\%) & $\mathcal{S}~(\sigma)$ \\ \hline
    $X(3960)$    & $0^{++}$ & $3956\pm5\pm10$ & $43\pm13\pm8$ & $25.4\pm7.7\pm5.0$ & 12.6 (14.6) \\
    $X_0(4140)$  & $0^{++}$ & $4133\pm6\pm6$ & $67\pm17\pm7$ & $16.7\pm4.7\pm3.9$ & 3.8 (4.1) \\
    $\psi(4260)$ & $1^{--}$ & 4230~\cite{PDG2018} & 55~\cite{PDG2018} & $3.6\pm0.4\pm3.2$ & 3.2 (3.6) \\
    $\psi(4660)$ & $1^{--}$ & 4633~\cite{PDG2020} & 64~\cite{PDG2020} & $2.2\pm0.2\pm0.8$ & 3.0 (3.2) \\
    NR & $0^{++}$ & - & - & $46.1\pm13.2\pm11.3$ & 3.1 (3.4) \\
    \hline
    \end{tabular}
\end{table}

The  $J^{PC}$ assignment for the system of a pair of oppositely-charged pseudoscalar mesons must be in the series $0^{++}$, $1^{--}$, $2^{++}$, etc. States with higher intrinsic spin are not expected to contribute significantly in the current dataset.
To determine the $X(3960)$ quantum numbers, fits with the baseline model are performed under alternative $J^{PC}$ hypotheses, $1^{--},~2^{++}$, instead of $0^{++}$. 
The significance to reject a $J^{PC}$ hypothesis is computed as $\sqrt{\Delta(-2\ln\Like)},$ where $\Delta(-2\ln\Like)=-(2\ln\Like(0^{++}) - 2\ln\Like(J^{PC}))$, and indicates the likelihood difference between the fits for the preferred $0^{++}$ assignment and for each alternative $J^{PC}$ hypothesis. 
To ensure that for different $J^{PC}$ hypotheses this resonance corresponds to the same particle, the mass and the width are limited to be within a $\pm3\,\sigma$ range of the baseline fit results. The $0^{++}$ assignment is preferred over $1^{--}$ and $2^{++}$ hypotheses by  $9.3\,\sigma$ and $12.3\,\sigma$, respectively. Similarly, replacing the baseline $0^{++}$ assignment by $1^{--}$ or $2^{++}$ for the $X_0(4140)$ state deteriorates the fit quality. 
The $0^{++}$ assignment is favoured over $1^{--}$ ($2^{++}$) hypothesis at a $3.5\,\sigma$ ($4.2\,\sigma$) level. Within the baseline model this $0^{++}$ state produces the dip around 4140~\mev via destructive interference with the $0^{++}$ NR  and $X(3960)$ components, 
with the interference fractions of, respectively, $(-22.4\pm6.4$)\% and ($-5.2\pm3.9$)\%, where the uncertainties are statistical only.

%% systematics
Systematic uncertainties on the measured resonance properties are evaluated, and are summarised in Table~\ref{tab:sys_dlz} in the supplemental material~\cite{sup}. 
Corrections, derived from calibration samples, are applied to account for possible discrepancies between data and simulation in the hardware trigger and particle-identification responses.
The uncertainty due to the limited size of the simulation samples is evaluated using the bootstrap method~\cite{efron:1979}. Additional resonances, not included in the baseline model (states in the \Dsp\Dsm system: $0^{++}$  $\chi_{c0}(4500)$ and $\chi_{c0}(4700)$~\cite{LHCb-PAPER-2020-044}, $1^{--}$ $\psi(4040)$, $\psi(4160)$ and $\psi(4415)$~\cite{PDG2020}, and $2^{++}$ $\chictwo(3930)$~\cite{LHCb-PAPER-2020-025}; and in the \Dsm\Kp system: $0^+$ $\Dbar^*_0(2300)^0$~\cite{PDG2020}, $1^-$ $\Dbar_1^{*}(2600)^0$~\cite{PDG2020,LHCb-PAPER-2019-027} and $\Dbar_1^{*}(2760)^0$~\cite{LHCb-PAPER-2015-007}, and $2^+$ $\Dbar_2^{*}(2460)^0$~\cite{PDG2020}) are utilised to estimate the uncertainty due to insufficient consideration of possible amplitude components. None of these states significantly improve the baseline model.
The \cquark\cquarkbar\squark\squarkbar candidates $\chiczero(4500)$ and $\chiczero(4700)$ have statistical significances of 0.8$\,\sigma$ and 1.3$\,\sigma$, respectively, and their fit fractions are $(0.6\pm1.0)\%$ ($<3.5\%$ at 90\% confidence level) and $(2.4\pm1.8)\%$ ($<6.7\%$ at 90\% confidence level), where the uncertainties are statistical.
The Blatt--Weisskopf hadron size is varied between 1.5 and 4.5~$\gev^{-1}$.
The fixed masses and widths of two baseline $\psi$ states are varied by their corresponding uncertainties. 
The Flatt\'{e}-like parameterisation for the $X(3960)$ state is replaced by a constant-width relativistic Breit--Wigner function.
The uncertainty due to the possible bias of the \sFit method is evaluated using pseudoexperiments.
The total systematic uncertainties on mass, width, and fit fraction are obtained by adding all contributions in quadrature, assuming that each source is independent. Regarding the total significance for each component in the baseline model, the smallest significance among these systematic tests is selected.

The measured mass and width of the $X(3960)$ state are consistent with those of the $\chiczero(3930)$ meson~\cite{LHCb-PAPER-2020-025} within $3\,\sigma$. 
Assuming that the $X(3960)$ in the $\Dsp\Dsm$ system and the $\chi_{c0}(3930)$ in the $\Dp\Dm$ system are the same state, the baseline model is extended by adding a second channel~(\Dp\Dm) in the Flatt\'{e}-like parameterisation.
The corresponding fit projections and numerical results can be found  in the supplemental material~\cite{sup}.
The likelihood is essentially unchanged 
while the n.d.f. is increased by one compared to the baseline fit. The coupling strength of the $X(3960)$ state to $\Dsp\Dsm$~($\Dp\Dm$) is found to be $0.33\pm1.18$~($0.15\pm0.33$)\gev.
The masses and fit fractions of all components are consistent with those in the baseline one-channel Flatt\'{e}-like model.

In the case that the $X(3960)$ and $\chi_{c0}(3930)$ states are the same particle, the partial width ratio of such an $X$ resonance decaying to \Dsp\Dsm and \Dp\Dm final states is calculated as
\begin{align}
\label{eq:ratio}
  {\Gamma(X\to\Dp\Dm)\over \Gamma({X\to\Dsp\Dsm})} = {\mathcal{B}^{(1)} \, \mathcal{F}_X^{(1)} \over \mathcal{B}^{(2)}  \, \mathcal{F}_X^{(2)}} = 0.29 \pm 0.09 \pm 0.10 \pm 0.08,
\end{align}
where the superscripts (1) and (2) indicate the \BuDDK and \BuDsDsK channels, respectively,  
$\mathcal{F}_X^{(1)}=(3.70\pm0.92)\%$ is the fit fraction of the $\chi_{c0}(3930)$ state in the \BuDDK decay~\cite{LHCb-PAPER-2020-025},  $\mathcal{F}_X^{(2)}$ is the fit fraction of the $X(3960)$ resonance presented in this Letter, and the branching fraction ratio $\mathcal{B}^{(1)}/\mathcal{B}^{(2)}$ is taken from the accompanying paper~\cite{LHCb-PAPER-2022-019}. 
The first uncertainty is statistical, the second systematic, and
the third is due to uncertainties in the measured branching fractions, \BrD and \BrDs~\cite{PDG2020}, and the uncertainty on $\mathcal{F}_X^{(1)}$~\cite{LHCb-PAPER-2020-025}.
This ratio is compatible with that of the couplings mentioned above.

It is well known that the creation of an \squark\squarkbar quark pair from the vacuum is suppressed relative to \uquark\uquarkbar or \dquark\dquarkbar pairs.
Moreover, the $X\to\Dsp\Dsm$ decay, occurring near the threshold, has a rather smaller phase-space factor than that of $X\to\Dp\Dm$. These two features indicate that $\Gamma(X\to\Dp\Dm)$ should be considerably larger than $\Gamma(X\to\Dsp\Dsm)$ if $X$ does not have any intrinsic \squark\squarkbar content. However, the value measured in Eq.~\eqref{eq:ratio} contradicts this expectation. This implies that the $X(3960)$ and $\chiczero(3930)$ are either not the same resonance, or they are the same non-conventional charmonium-like state, for instance, a candidate containing the dominant \cquark\cquarkbar\squark\squarkbar constituents predicted in recent theoretical  models~\cite{3915:Lebed,3915:2017dpy,3915:LQCD,3915:enwang,3915:jlping,Meng:2020cbk,Dong:2021juy}. Further studies are needed to gain insights into the nature of the $\Dsp\Dsm$ threshold enhancement, in particular the measurement of the relative branching fraction for the $D_{(s)}\overline{D}_{(s)}$ and $\omega\jpsi$ channels, produced in a different environment, such as from two-photon fusion processes by the \belle II experiment.

There is no obvious candidate within  conventional charmonium multiplets for $X(3960)$ or $\chi_{c0}(3930)$ assignment.
First of all, the mass of the $\chi_{c0}(3930)$ state is far from predictions for the $\chi_{c0}(3P)$,  which lies within the range 4131\textendash4292\mev~\cite{Barnes:2005pb,cc:KTChao}.
For the $\chi_{c0}(2P)$ state, most potential models predict a mass in the range 3842\textendash3868\mev~\cite{Radford:2007vd,cc:KTChao,3Pj:Ping}, except the Godfrey-Isgur model which gives 3916\mev ~\cite{Barnes:2005pb}. Second, the $\chi_{c0}(3930)$ state interpreted as $\chi_{c0}(2P)$  would give too small a mass splitting with respect to the $\chi_{c2}(3930)$ state~\cite{LHCb-PAPER-2020-025} identified as $\chi_{c2}(2P)$~\cite{Barnes:2005pb,Radford:2007vd,cc:KTChao,3Pj:Ping}.
In addition, interpreting the $\chiczero(3930)$ state as the $\chiczero(2P)$ charmonium would result in inconsistent decay widths, as the Okubo-Zweig-Iizuka (OZI)~\cite{Okubo:1963fa,Iizuka:1966fk} suppressed channel $\chiczero(3930)\to\omega\jpsi$ has a decay width larger than theoretical expectations, whereas the $S$-wave OZI-allowed $\chiczero(3930)\to\D\Dbar$ mode has smaller decay width than the expectations~\cite{3Pj:Ping,3860:Guo}.
As a consequence, neither the $X(3960)$ nor the $\chiczero(3930)$ is likely to be a pure $\chiczero(2P)$ or $\chiczero(3P)$ charmonium state.

To test the possibility that the dip in the $\Dsp\Dsm$ invariant mass around 4140\mev can be produced by the opening of the nearby $\jpsi\phi$ threshold, without introducing an additional resonance, we employ a simple $K$-matrix model that contains the single  resonance $X(3960)$ and two coupled channels, $\Dsp\Dsm$ and $\jpsi\phi$.
The $K$-matrix reads

\begin{equation}
\label{eq:kmatrix}
\left(\begin{matrix}
\mathcal{K}_{\Dsp\Dsm\to\Dsp\Dsm} & \mathcal{K}_{\Dsp\Dsm\to\jpsi\phi}  \\ 
\mathcal{K}_{\jpsi\phi\to\Dsp\Dsm} & \mathcal{K}_{\jpsi\phi\to\jpsi\phi}
\end{matrix}\right)
\equiv
\left(\begin{matrix}
\mathcal{K}_{11} & \mathcal{K}_{12}  \\ 
\mathcal{K}_{21} & \mathcal{K}_{22}
\end{matrix}\right),
\end{equation}
where $\mathcal{K}_{12}=\mathcal{K}_{21}$, and the subscripts $1$ and $2$ represent $\Dsp\Dsm$ and $\jpsi\phi$ final states, respectively. 
One possible choice for the $2\times2$ $K$-matrix parameterisation~\cite{PDG2020} is
\begin{equation}
\label{eq:Kba}
\mathcal{K}_{ba}(m)=\sum_{R}\frac{g^R_b g^R_a}{M^2_R-m^2}+f_{ba},
\end{equation}
where $M_R$ refers to the bare mass of the resonance $R$, $m$ is the $\Dsp\Dsm$ invariant mass, $g^R_a$ denotes the bare coupling of the resonance $R$ to the channel $a$, and the $f_{ba}$ is a real matrix parameterising the non-pole part of the $K$-matrix. As the $X(3960)$ mass is about 160\mev lower than the $\jpsi\phi$ threshold and its width is less than $50$\mev, the coupling of the $X(3960)$ state to $\jpsi\phi$ should be negligible, giving $g^R_2=0$. This results in the $X(3960)$ resonance entering the $\mathcal{K}_{11}$ element only. The production amplitude is expressed in the $P$-vector formalism~\cite{PDG2020,kmatrix:1972,kmatrix:2015}, which gives
\begin{equation}
{\cal P}_b(m)=\sum_{R}\frac{\beta_{R} g^R_b}{M^2_R-m^2}+\beta_{b},
\end{equation}
where $\beta_R$ and $\beta_b$ are complex free parameters due to rescattering effects or missing channels~\cite{PDG2018}.
The amplitude $\mathcal{M}$ is
\begin{equation}
\label{eq:amp_kmatrix}
\mathcal{M}_{a}=\sum_b(I-i\rho\mathcal{K})^{-1}_{ab}\mathcal{P}_{b},
\end{equation}
where $\rho={\rm diag} \{\rho_{11}, \rho_{22}\}$ is the diagonal matrix composed of phase-space factors, $I$ represents the identity matrix, and $a=1$ for the \Dsp\Dsm channel under consideration.

The fit demonstrates that the dip around the $\jpsi\phi$ threshold can also be modelled by the $\jpsi \phi \to \Dsp\Dsm$  rescattering, and results in a $-2\ln\Like$ that is worse by 6.0, while the n.d.f. is increased by one, compared to the baseline fit.
The fit projections and numerical results can be found in the supplemental material~\cite{sup}.
Since the fit quality of the $K$-matrix parameterisation is close to that of the baseline model, a strong conclusion cannot be drawn whether the dip is due to destructive interference with the $X_0(4140)$ resonance or caused by the $\jpsi\phi\to\Dsp\Dsm$ rescattering. 

In addition, it is found that the fits with the two-channel Flatt\'e-like and $K$-matrix parameterisations are unstable, due to having too many free parameters for the limited data sample size.
Consequently, the statistical uncertainties for some  parameters are large. 
Therefore, neither of these parameterisations are taken as the baseline model.

In conclusion, the first amplitude analysis of the \BuDsDsK decay is performed using $pp$ collision data with an integrated luminosity of 9\invfb collected by the LHCb experiment between 2011 and 2018. 
A peaking structure near the $\Dsp\Dsm$ mass threshold, denoted as $X(3960)$, is observed with a significance larger than 12$\,\sigma$. Its quantum numbers are  determined to be $J^{PC}=0^{++}$, favoured over $1^{--}$ or $2^{++}$ with a significance greater than 9$\,\sigma$. 
As argued above, the $X(3960)$ and $\chiczero(3930)$ states are unlikely to be the same pure conventional charmonium state.
The $X(3960)$ resonance presented in this Letter could be a candidate for an exotic state predominantly consisting of  \cquark\cquarkbar\squark\squarkbar constituents, as suggested in recent theoretical literature~\cite{3915:Lebed,3915:2017dpy,3915:LQCD,Meng:2020cbk,Dong:2021juy,3915:enwang,3915:jlping}.
If predominant \cquark\cquarkbar\squark\squarkbar content is confirmed, this state should be labelled $T_{\psi\phi}^f(3960)$ in the new naming scheme for exotic hadrons~\cite{LHCb-PUB-2022-013}.
In addition, a dip around 4140\mev can be described either by a $0^{++}$ $X_0(4140)$ resonance having a significance of 3.5$\,\sigma$, or the coupled-channel effect of the $\jpsi\phi\leftrightarrow\Dsp\Dsm$  reaction. 
The data from the forthcoming Run 3 of the LHCb experiment and from the Belle II experiment will be critical to clarify the nature of these phenomena.

%% file: supplementary.tex
\clearpage

\section*{Supplemental material}
\label{sec:supplementary}
%\onecolumngrid
\setcounter{equation}{0}
\setcounter{figure}{0}
\setcounter{table}{0}
\setcounter{section}{0}
%\setcounter{page}{1}
%\makeatletter
\renewcommand{\theequation}{S\arabic{equation}}
\renewcommand{\thefigure}{S\arabic{figure}}
\renewcommand{\thetable}{S\arabic{table}}
\newcommand\ptwiddle[1]{\mathord{\mathop{#1}\limits^{\scriptscriptstyle(\sim)}}}

% \begin{table}[htb]
% \hspace{-0.8cm}
% \vspace{0.2cm}
% \centering
% \renewcommand\arraystretch{1.1} %spread row spacing 1.1 times than default
% \caption{Summary of the systematic uncertainties on the measured properties in the Dalitz plot analysis.}
% \label{tab:sys_dlz}
% \begin{tabular}{l|c|cccccccc}\hline\hline
% \multicolumn{2}{l|}{Source} & L0 & MC & PID & Comp. & Bl-W & $M_0\&\Gamma_0$ & Model & Total \\ \hline
% \multirow{3}*{$X(3960)$}
% & $M_0$ (MeV)       & 0 & 2 & 0  & 2 & 0 & 1 & 11 & 11 \\
% & $\Gamma_0$ (MeV)  & 0 & 1 & 0  & 3 & 1 & 2 & 9 & 10 \\
% & $\mathcal{F}$ (\%)         & 0.6 & 0.7 & 0.5 & 7.1 & 0.0 & 2.8 & 1.0 & 7.8\\ 

% \multirow{3}*{$X_0(4140)$}
% & $M_0$ (MeV)      & 0 & 1 & 0 & 10 & 1 & 4 & 1 & 11\\
% & $\Gamma_0$ (MeV) & 0 & 1 & 2 & 5  & 1 & 4 & 1 & 7\\
% & $\mathcal{F}$ (\%)        & 0.1 & 0.5 & 0.0 & 6.9 & 0.1 & 2.9 & 1.9 & 7.5 \\

% $\psi(4260)$
% & $\mathcal{F}$ (\%)        & 0.0 & 0.0 & 0.0 & 3.0 & 0.0 & 0.1 & 0.1 & 3.0\\

% $\psi(4660)$
% & $\mathcal{F}$ (\%)        & 0.0 & 0.0 & 0.0 & 0.4 & 0.0 & 0.1 & 0.2 & 0.4 \\

% NR
% & $\mathcal{F}$ (\%)        & 0.7 & 1.7 & 0.7 & 9.8 & 0.1 & 3.7 & 3.2 & 10.7 \\

% \hline\hline
% \end{tabular}
% \end{table}

\begin{table}[htb]
\centering
\renewcommand\arraystretch{1.1} %spread row spacing 1.1 times than default
\caption{Summary of the systematic uncertainties on the parameters measured in the amplitude analysis. Mass ($M_0$) and width ($\Gamma_0$) are in units of MeV, while fit fraction ($\mathcal{F}$) is in \%.}
\label{tab:sys_dlz}
\begin{tabular}{lccccccccc}\hline
         & \multicolumn{3}{c}{$X(3960)$} & \multicolumn{3}{c}{$X_0(4140)$} & $\psi(4260)$ & $\psi(4660)$ & NR \\ \cline{2-4} \cline{5-7} \cline{8-10}
Source   & $M_0$ & $\Gamma_0$ & $\mathcal{F}$ & $M_0$ & $\Gamma_0$ & $\mathcal{F}$ & $\mathcal{F}$ & $\mathcal{F}$ & $\mathcal{F}$ \\ \hline
Trigger          & 0   & 0   & 0.6   & 0   & 0   & 0.1   & 0.0   & 0.0   & 0.7   \\
Simulation sample size    & 2   & 1   & 0.7   & 1   & 1   & 0.5   & 0.0   & 0.0   & 1.7   \\
Particle identification  & 0   & 0   & 0.5   & 0   & 2   & 0.0   & 0.0   & 0.0   & 0.7   \\
Additional fit components       & 1   & 3   & 3.4   & 3  & 5   & 2.5   & 3.2   & 0.7   & 10.1   \\
Hadron size      & 0   & 1   & 0.0   & 1   & 1   & 0.1   & 0.0   & 0.0   & 0.1   \\
Fixed parameters & 1   & 2   & 2.8   & 4   & 4   & 2.9   & 0.1   & 0.1   & 3.7   \\
$X(3960)$ model  & 10  & 7   & 1.6   & 0   & 1   & 0.7   & 0.0   & 0.0   & 2.1   \\
\sFit bias       & 1.9 & 1.5 & 1.5   & 2.6 & 1.1 & 0.4   & 0.3   & 0.3   & 2.1   \\ \hline
Total            & 10  & 8  & 5.0   & 6  & 7   & 3.9   & 3.2   & 0.8   & 11.3  \\
\hline
\end{tabular}
\end{table}

\begin{table}[htb]
%\vspace{0.2cm}
%\hspace{-0.4cm}
\centering
\renewcommand\arraystretch{1.1} %spread row spacing 1.1 times than default
\caption{Main results found in two-channel Flatt\'{e}-like parameterisation, where the coupling strength of the $X(3960)$ state to $\Dsp\Dsm$ ($\Dp\Dm$) is obtained to be $0.33\pm1.18$ ($0.15\pm0.33$)\gev. Uncertainties are statistical only. The large uncertainty on $\Gamma_0$ for the $X(3960)$ state is due to the large uncertainty on the coupling strengths.}
\label{tab:twoflatte}
\begin{tabular}{lccccc}\hline
Contribution & $J^{PC}$ & $M_0$ (MeV) & $\Gamma_0$ (MeV) & $\mathcal{F}$ (\%) \\ \hline
$X(3960)$  & $0^{++}$ & $3951\pm14$ & $38\pm104$ & $25.0\pm7.6$\\
$X_0(4140)$  & $0^{++}$ & $4133\pm7$ & $67\pm16$ & $16.7\pm4.6$\\
$\psi(4260)$ & $1^{--}$ & 4230~\cite{PDG2018} & 55~\cite{PDG2018} & $3.6\pm0.4$ \\
$\psi(4660)$ & $1^{--}$ & 4633~\cite{PDG2020} & 64~\cite{PDG2020} & $2.2\pm0.2$ \\
NR & $0^{++}$ & - & - & $45.9\pm10.9$  \\
\hline
\end{tabular}
\end{table}

\begin{table}[htb]
\vspace{0.2cm}
\hspace{0.4cm}
\centering
\renewcommand\arraystretch{1.1} %spread row spacing 1.1 times than default
\caption{Main results found from the $K$-matrix fit. Uncertainties are statistical only. }
\label{tab:kmatrix}
\begin{tabular}{lcccccc}\hline
Contribution & $J^{PC}$ & $M_R$ (MeV) & $g^R_1$ (MeV) & $\Gamma_{0}$ (MeV) & $\mathcal{F}$ (\%) \\ \hline
$|\mathcal{M}_1|^2$  & $0^{++}$ & $3957\pm14$ & $1350\pm344$ & & $94.7\pm0.4$\\
$\psi(4260)$ & $1^{--}$ & 4230~\cite{PDG2018} & & 55~\cite{PDG2018} & $3.2\pm0.5$ \\
$\psi(4660)$ & $1^{--}$ & 4633~\cite{PDG2020} & & 64~\cite{PDG2020} & $2.1\pm0.2$ \\ \hline
$\beta_R$ &  \multicolumn{2}{c}{$(1,0i)$} & $\beta_1$ & \multicolumn{2}{c}{$(-1.2,2.5i)\pm(4.5,3.1i)$} \\ 
$\beta_2$ & \multicolumn{2}{c}{$(-137.2,-1.5i)\pm(2.7,218.6i)$} & $f_{11}$ & \multicolumn{2}{c}{$0.8\pm1.2$} \\
$f_{12}=f_{21}$ & \multicolumn{2}{c}{$0.1\pm0.1$} & $f_{22}$ & \multicolumn{2}{c}{$8.0\pm5.1$} \\
\hline
\end{tabular}
\end{table}

\begin{figure}[htbp]
\centering
%\hspace{-1.0cm}
\makebox[1.0\textwidth][c]{
\includegraphics[width=0.5\textwidth]{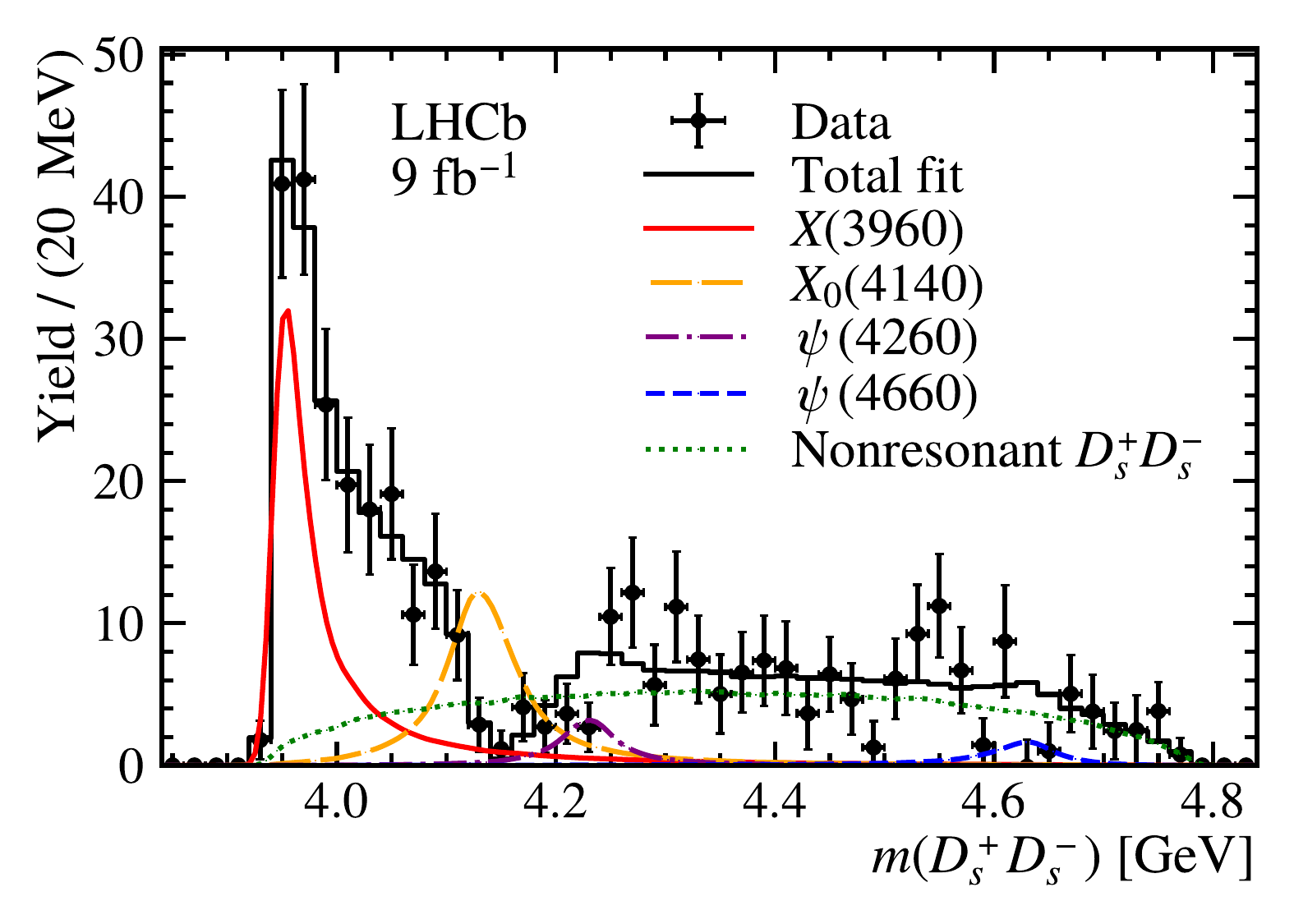}
\includegraphics[width=0.5\textwidth]{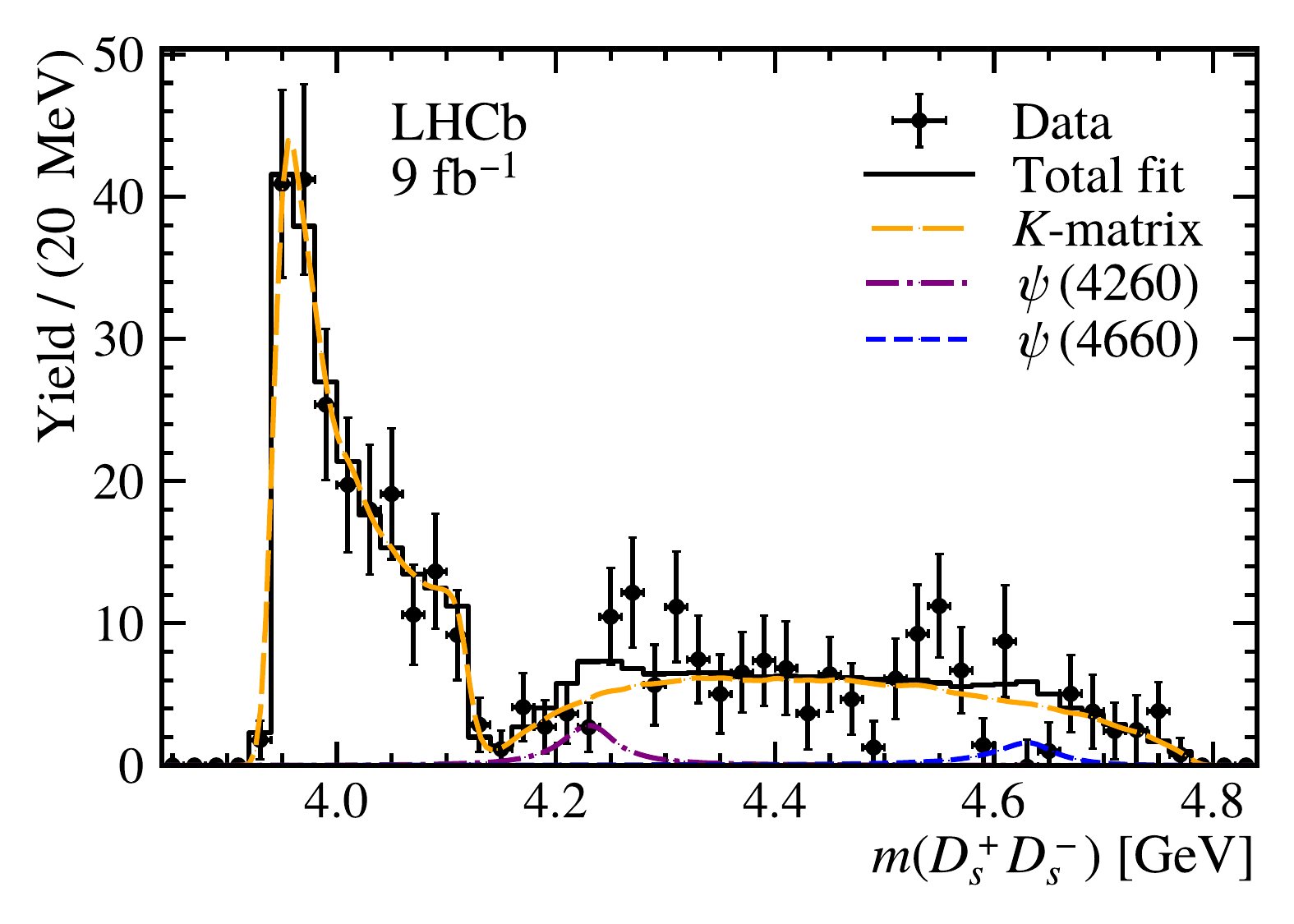} 
}
\caption{Background-subtracted distributions of the $\Dsp\Dsm$ invariant mass of $\Bp\to\Dsp\Dsm\Kp$ decays with fit results obtained from (left) two-channel Flatt\'{e}-like and (right) $K$-matrix parameterisations. The fit projections for the \Dsm\Kp and \Dsp\Kp invariant-mass spectra look very similar to the baseline model.
\label{fig:flatte-kmatrix}} 
\end{figure}

\clearpage

%% file: acknowledgements.tex
\section*{Acknowledgements}
%
% These Acknowledgements valid from 3-May-2019
%
\noindent We express our gratitude to our colleagues in the CERN
accelerator departments for the excellent performance of the LHC. We
thank the technical and administrative staff at the LHCb
institutes.
We acknowledge support from CERN and from the national agencies:
CAPES, CNPq, FAPERJ and FINEP (Brazil); 
MOST and NSFC (China); 
CNRS/IN2P3 (France); 
BMBF, DFG and MPG (Germany); 
INFN (Italy); 
NWO (Netherlands); 
MNiSW and NCN (Poland); 
MEN/IFA (Romania); 
%MSHE (Russia); 
MICINN (Spain); 
SNSF and SER (Switzerland); 
NASU (Ukraine); 
STFC (United Kingdom); 
DOE NP and NSF (USA).
We acknowledge the computing resources that are provided by CERN, IN2P3
(France), KIT and DESY (Germany), INFN (Italy), SURF (Netherlands),
PIC (Spain), GridPP (United Kingdom), 
%RRCKI and Yandex LLC (Russia), 
CSCS (Switzerland), IFIN-HH (Romania), CBPF (Brazil),
Polish WLCG  (Poland) and NERSC (USA).
We are indebted to the communities behind the multiple open-source
software packages on which we depend.
Individual groups or members have received support from
ARC and ARDC (Australia);
Minciencias (Colombia);
AvH Foundation (Germany);
EPLANET, Marie Sk\l{}odowska-Curie Actions and ERC (European Union);
A*MIDEX, ANR, IPhU and Labex P2IO, and R\'{e}gion Auvergne-Rh\^{o}ne-Alpes (France);
Key Research Program of Frontier Sciences of CAS, CAS PIFI, CAS CCEPP, 
Fundamental Research Funds for the Central Universities, 
and Sci. \& Tech. Program of Guangzhou (China);
%Key Research Program of Frontier Sciences of CAS, CAS PIFI,
%Thousand Talents Program, and Sci. \& Tech. Program of Guangzhou (China);
%RFBR, RSF and Yandex LLC (Russia);
GVA, XuntaGal, GENCAT and Prog.~Atracci\'on Talento, CM (Spain);
SRC (Sweden);
the Leverhulme Trust, the Royal Society
 and UKRI (United Kingdom).

%% file: Authorship_LHCb-PAPER-2022-018.tex
% LHCb collaboration author list
% Data extracted on April 28th, 2023 at 11:11am for paper reference LHCb-PAPER-2022-018
\centerline
{\large\bf LHCb collaboration}
\begin
{flushleft}
\small
R.~Aaij$^{32}$\lhcborcid{0000-0003-0533-1952},
A.S.W.~Abdelmotteleb$^{50}$\lhcborcid{0000-0001-7905-0542},
C.~Abellan~Beteta$^{44}$,
F.~Abudin{\'e}n$^{50}$\lhcborcid{0000-0002-6737-3528},
T.~Ackernley$^{54}$\lhcborcid{0000-0002-5951-3498},
B.~Adeva$^{40}$\lhcborcid{0000-0001-9756-3712},
M.~Adinolfi$^{48}$\lhcborcid{0000-0002-1326-1264},
H.~Afsharnia$^{9}$,
C.~Agapopoulou$^{13}$\lhcborcid{0000-0002-2368-0147},
C.A.~Aidala$^{77}$\lhcborcid{0000-0001-9540-4988}
% ...

{\footnotesize \it

% ...
$^{9}$Universit{\'e} Clermont Auvergne, CNRS/IN2P3, LPC, Clermont-Ferrand, France\\
% ...
$^{13}$LPNHE, Sorbonne Universit{\'e}, Paris Diderot Sorbonne Paris Cit{\'e}, CNRS/IN2P3, Paris, France\\
% ...
$^{32}$Nikhef National Institute for Subatomic Physics, Amsterdam, Netherlands\\
% ...
$^{40}$Instituto Galego de F{\'\i}sica de Altas Enerx{\'\i}as (IGFAE), Universidade de Santiago de Compostela, Santiago de Compostela, Spain\\
% ...
$^{44}$Physik-Institut, Universit{\"a}t Z{\"u}rich, Z{\"u}rich, Switzerland\\
% ...
$^{48}$H.H. Wills Physics Laboratory, University of Bristol, Bristol, United Kingdom\\
% ...
$^{50}$Department of Physics, University of Warwick, Coventry, United Kingdom\\
% ...
$^{54}$Oliver Lodge Laboratory, University of Liverpool, Liverpool, United Kingdom\\
% ...
$^{62}$Syracuse University, Syracuse, NY, United States\\
% ...
$^{77}$University of Michigan, Ann Arbor, MI, United States, associated to $^{62}$\\
\bigskip
% ...
}
\end{flushleft}